%% file: main.tex
\documentclass[a4paper,twoside]{article}

\usepackage{epsfig}
\usepackage[export]{adjustbox}
\usepackage{subcaption}
\usepackage{calc}
\usepackage{amssymb}
\usepackage{amstext}
\usepackage{amsmath}
\usepackage{amsthm}
\usepackage{multicol}
\usepackage{pslatex}
\usepackage{apalike}
\usepackage[bottom]{footmisc}
\usepackage{url}
\usepackage{hyperref}
\usepackage{xcolor, soul}
\usepackage{graphicx}
\usepackage{epstopdf}
\sethlcolor{yellow}
\usepackage{multirow}
\usepackage{rotating}
\usepackage{framed}
\usepackage{SCITEPRESS}     % Please add other packages that you may need BEFORE the SCITEPRESS.sty package.

\hypersetup{
    colorlinks = true,
    urlcolor   = blue,
    linkcolor  = blue,
    citecolor  = blue,
    filecolor  = blue
}

\begin{document}

\title{BIP! NDR (NoDoiRefs): A Dataset of Citations From Papers Without DOIs in Computer Science Conferences and Workshops}

\author{\authorname{Paris Koloveas\sup{1,2}\orcidAuthor{0000-0003-2376-089X}, Serafeim Chatzopoulos\sup{2}\orcidAuthor{0000-0003-1714-5225}, Christos Tryfonopoulos\sup{1}\orcidAuthor{0000-0003-0640-9088} and Thanasis Vergoulis\sup{2}\orcidAuthor{0000-0003-0555-4128}}
\affiliation{\sup{1}University of the Peloponnese, Tripolis, Greece}
\affiliation{\sup{2}IMSI, Athena RC, Athens, Greece}
\email{\{pkoloveas, trifon\}@uop.gr, \{schatz, vergoulis\}@athenarc.gr}}


\input{sections/abstract.tex}

\onecolumn \maketitle \normalsize \setcounter{footnote}{0} \vfill

\input{sections/sec1-introduction.tex}
\input{sections/sec3-methodology.tex}
\input{sections/sec4-dataset_structure-stats.tex}
\input{sections/sec5-conclusions.tex}

\bibliographystyle{apalike}
\small
\bibliography{refs}

\end{document}

%% file: sections/abstract.tex
\abstract{In the field of Computer Science, conference and workshop papers serve as important contributions, carrying substantial weight in research assessment processes, compared to other disciplines. However, a considerable number of these papers are not assigned a Digital Object Identifier (DOI), hence their citations are not reported in widely used citation datasets like OpenCitations and Crossref, raising limitations to citation analysis. While the Microsoft Academic Graph (MAG) previously addressed this issue by providing substantial coverage, its discontinuation 
has created a void in available data. 
BIP! NDR aims to alleviate this issue and enhance the research assessment processes within the field of Computer Science. To accomplish this, it leverages a workflow that identifies and retrieves Open Science papers lacking DOIs from the DBLP Corpus, and by performing text analysis, it extracts citation information directly from their full text. The current version of the dataset contains more than 510K citations made by approximately 60K open access Computer Science conference or workshop papers that, according to DBLP, do not have a DOI. }

%% file: sections/sec1-introduction.tex
\section{Introduction}\label{sec:introduction}

A \emph{(bibliographic) citation} refers to a conceptual (directional) link that connects a research work (usually a publication) which contains a reference to (i.e., ``cites'') another work (which is being ``cited''). During the last decades, citations have become one of the most important types of bibliographic metadata~\cite{peroni-qss}. The main reason for that is that they are often considered as proxies of scientific impact, since a citation can be interpreted as an acknowledgement for the contribution of the cited work into the citing one (although this might not always be the case~\cite{YTN2017,AER2013}). As a result, they have been instrumental in scientometrics, becoming the basis for the calculation of various research impact indicators~\cite{bipdb}. Such indicators have been used to facilitate scientific knowledge discovery (e.g., they have been used by academic search engines to help researchers prioritise their reading~\cite{bip-finder}), monitor research production~\cite{observatory}, assist research assessment processes, and in many other applications.

Various sources of citation data have become available during the previous decades to address the needs of use-cases like the aforementioned ones. Apart from proprietary and restrictive sources, like Clarivate Analytics' Web of Science, Google Scholar and the Microsoft Academic Graph (MAG)~\cite{mag-qss}, due to the raised popularity of the Open Science movement, a couple of open datasets that provide citations (e.g., OpenCitations\footnote{OpenCitations: \href{https://opencitations.net}{opencitations.net}}, the OpenAIRE Graph\footnote{OpenAIRE Graph: \href{https://graph.openaire.eu}{graph.openaire.eu}}) have also become available during the last years. 
Almost all of them report citations as DOI-to-DOI pairs, failing to cover citations that involve publications  for which a DOI has not been assigned. This may not be a significant problem for many disciplines, but in Computer Science, a considerable number of conferences and workshops do not assign DOIs to their papers. In addition, in this field, conference and workshop papers are peer reviewed and, historically, serve as important contributions, carrying significant weight in research assessment processes. As a result, if they are not considered during citation analyses, this can overlook an important part of scientific production and even introduce bias. In the past, Microsoft Academic Graph (MAG) was partially covering this gap by also offering citations for papers that do not have a DOI.  However, since its discontinuation in December $2021$, this data collection is no longer maintained and updated, thus its coverage is continuously declining.

In this work, we introduce BIP! NDR, an open dataset that aims to cover this gap, improving research assessment processes and other relevant applications within the field of Computer Science. The dataset is constructed based on a workflow that identifies and retrieves Open Science publications lacking DOIs from DBLP\footnote{DBLP: \url{https://dblp.uni-trier.de/}}, the most widely known bibliographic database for publications from Computer Science, and then performs text analysis to extract citation information directly from the respective manuscripts. The current version of the dataset contains more than $510$K citations made by approximately $60$K Computer Science conference or workshop papers that,
according to DBLP, do not have a DOI. We plan to frequently update the dataset so that it can become an important resource for citations in Computer Science that are missing from the most important citation datasets. This is a valuable addition to the toolboxes of scientometricians so that they can perform more concrete analysis in the Computer Science domain. 

\noindent \textbf{Outline.} The rest of the manuscript is organized as follows: in Section~\ref{sec3-methodology} we elaborate on the technical details related to the production of the BIP! NDR dataset; in Section~\ref{sec4-dataset-structure-stats} we discuss the structure of the dataset; finally, in Section~\ref{sec5-conclusions} we conclude the work while also discussing future planned extensions.

%% file: sections/sec3-methodology.tex
\section{Dataset Production Workflow}\label{sec3-methodology}

\begin{figure*}
    \centering
    \includegraphics[width=\textwidth]{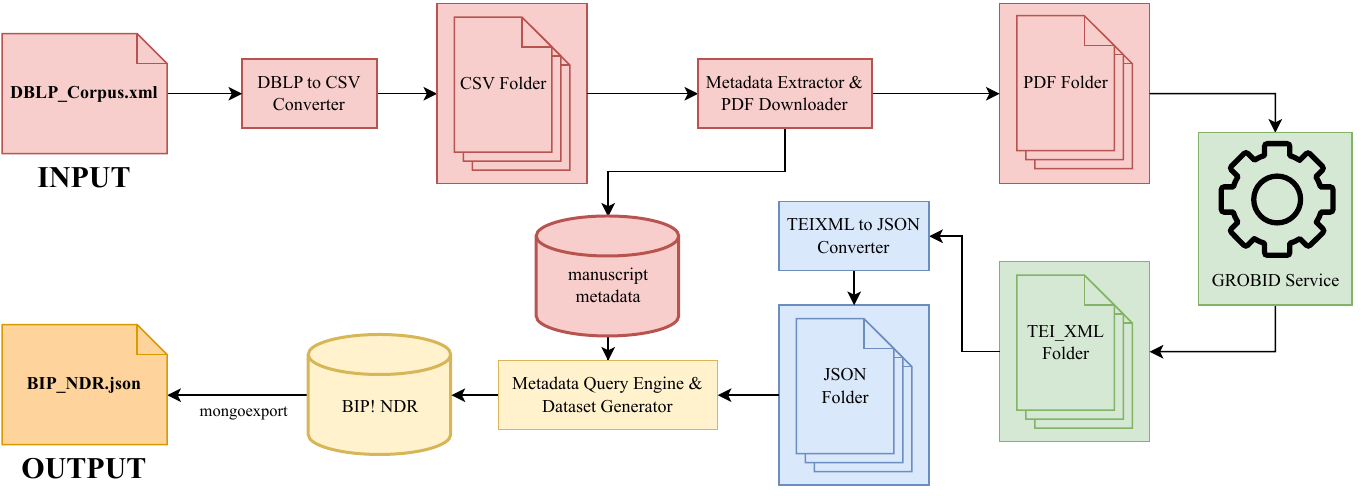}
    \caption{A high-level overview of the dataset production workflow.}\label{fig1:pipeline-overview}
\end{figure*}

In this section, we discuss the BIP! NDR dataset production workflow and we elaborate on the technical details of its various components. 
The source code of the production workflow is available as open source on GitHub\footnote{BIP! NDR: \href{https://github.com/athenarc/bip-ndr-workflow}{github.com/athenarc/bip-ndr-workflow}}.
A high-level overview of the workflow is depicted in Figure~\ref{fig1:pipeline-overview}.

The main input to the workflow is the DBLP Corpus, which we use to collect URLs hosting Open Access manuscripts from the field of Computer Science, focusing on those that do not have a DOI. We collect these manuscripts so that we will be able to extract citations from the respective PDF files. 
DBLP~\cite{ley2002dblp,ley2009dblp} consolidates scholarly metadata from several open sources which cover the Computer Science field and is largely manually curated and frequently updated. 

As a result, this collection is ideal for our purposes. Our analysis shows that out of the approximately 320K open access conference publications, approximately 260K do not have a DOI. These publications are the ones that we aim to cover through the evolution of our dataset.
The current version of our dataset (v0.1)~\cite{Koloveas-bipndr} is based on the November 2022 Monthly Snapshot of DBLP~\cite{dblp-dataset}. The DBLP Corpus comes in XML format with all the bibliographic entries together in a single file. Therefore, as a first step, we use \texttt{dblp-to-csv}\footnote{dblp-to-csv: \href{https://github.com/ThomHurks/dblp-to-csv}{github.com/ThomHurks/dblp-to-csv}} to split the corpus into separate CSV files, grouped by publication type. We further process these CSV files to (a) extract manuscript metadata and store them in a document-oriented database, and (b) follow the included links to download the PDF files of Open Access papers. These operations ensure that the structured manuscript metadata from the DBLP Corpus are easily accessible to our workflow for querying and further processing.

For the next step of our workflow, we needed a tool to extract information from the PDF files while maintaining the headers, structure and sectioning of the manuscript. 
After a thorough evaluation of the literature regarding the tools used for reference extraction from PDFs, we concluded that based both on surveys~\cite{TCS2018,MJS2023}, and prominent works that required extensive bibliography parsing~\cite{LWN2020,NML2021}, GROBID~\cite{L2009} is currently the best tool for the task. GROBID converts the PDF files to the TEI XML publication format\footnote{TEI XML format: \href{https://tei-c.org/release/doc/tei-p5-doc/en/html/SG.html}{tei-c.org/release/doc/tei-p5-doc/en/html/SG.html}}. Apart from the PDF extraction capabilities, GROBID offers a consolidation option to resolve extracted bibliographical references using services like biblio-glutton\footnote{biblio-glutton: \href{https://github.com/kermitt2/biblio-glutton}{github.com/kermitt2/biblio-glutton}} or the CrossRef REST API\footnote{Crossref API: 
\href{https://www.crossref.org/documentation/retrieve-metadata/rest-api/}{crossref.org/documentation/retrieve-metadata/rest-api/}}.
We apply this consolidation option to our workflow, and GROBID sends a request to the Crossref web service~\cite{crossref} for each extracted citation. If a core of metadata (such as the main title and the first author) is correctly identified, the system retrieves the full publisher's metadata. These metadata are then used for correcting the extracted fields and for enriching the results. We utilize this output to potentially identify the DOI of a publication and attempt to match it with a DBLP entry.

The TEI XML files that GROBID produces are useful for identifying the structure of a manuscript, but are very verbose and are not convenient to process in large volumes. For that reason, we have created a \emph{TEIXML to JSON Converter} that turns the files into JSON format. This conversion process involves extracting relevant information from the TEI XML files and mapping it to the corresponding JSON structure. The resulting JSON files are smaller in size and are compatible with a wide range of tools for processing.

At this point, we have reached the core functionality of our workflow, the process of \emph{querying the DBLP metadata} for the bibliographic references of the papers in our collection. This process queries the manuscript metadata database for each document in the JSON Folder. For each document, we parse the reference list and we first check if a DOI exists in a publication entry. If it exists, we query our database based on the DOI. If a result is returned, we store the \texttt{dblp\_id}, the \texttt{doi}, as well as, the \texttt{bibliographic\_reference} extracted from the JSON file. Otherwise, we query based on the publication title. On a positive result, we store the previously mentioned fields to the dataset entry. If neither the publication title nor the DOI return a positive result, the publication could not be found in our DBLP metadata, so we store only the \texttt{doi} and \texttt{bibliographic\_reference} from the JSON file. This process ultimately creates the ``BIP! NDR'' collection which constitutes our dataset.

The final step involved using the \emph{mongoexport} utility to export the ``BIP! NDR'' collection from MongoDB into the final JSONL file.
The exported file served as the culmination of the dataset generation process, providing a structured collection of scholarly data ready for research and analysis.

%% file: sections/sec4-dataset_structure-stats.tex
\section{The BIP! NDR Dataset}\label{sec4-dataset-structure-stats}

\begin{figure*}[!t]
\centering

\begin{framed}
    \centering
    \begin{verbatim}
{
   "_id":{
      "$oid": "6460a56bda929a01210c1b57"
   },
   "citing_paper": {
      "dblp_id": "conf/ecsa/GasperisPF21"
   },
    "cited_papers": [
      {
         "dblp_id": "journals/sigpro/AlbusacCLVL09",
         "doi": "10.1016/j.sigpro.2009.04.008",
         "bibliographic_reference": "J. Albusac, J. Castro-Schez, L. Lopez-Lopez, 
            D. Vallejo, L. Jimenez-Linares, A supervised  learning approach to 
            automate the acquisition of knowledge in surveillance systems, 
            Signal Processing 89 (2009) 2400-2414. 
            doi:https://doi.org/10.1016/j.sigpro.2009.04.008, special Section: 
            Visual Information Analysis for Security."
      },
      {
         "dblp_id": "journals/cssp/Elhoseny20",
         "doi": "10.1007/s00034-019-01234-7",
         "bibliographic_reference": "M. Elhoseny, Multi-object detection and tracking 
             (modt) machine learning model for real-time video surveillance 
             systems, Circuits, Systems, and Signal Processing 39 (2020) 611-630. 
             doi:10.1007/s00034-019-01234-7."
      },
      {
         "doi":"10.23919/IRS.2019.8768102",
         "bibliographic_reference":"F. Opitz, K. Dästner, B. Roseneckh-Köhler, 
             E. Schmid, Data analytics and machine learning in wide area surveillance 
             systems, in: 2019 20th International Radar Symposium (IRS), 2019, 
             pp. 1-10. doi:10.23919/IRS.2019.8768102."
      },
      {
         "dblp_id":"journals/rfc/rfc3411",
         "bibliographic_reference":"D. Harrington, R. Presuhn, B. Wijnen, 
            An architecture for describing simple network management protocol (snmp) 
            management frameworks, 2002. doi:10.17487/RFC3411."
      }
    ]
}
     \end{verbatim}
\end{framed}
\caption{Data structure of the BIP! NDR dataset.}\label{tab:dataset-structure}
\end{figure*}

In this section we present the structure of the dataset along with some basic statistics of the current version. 
The dataset is formatted as a JSON Lines (JSONL)\footnote{JSON Lines data format: \href{https://jsonlines.org/}{jsonlines.org}} file where each line contains a valid JSON object. This file format 
enables file splitting and data streaming as the dataset grows in size. 
An indicative record (in JSON format) of the BIP! NDR dataset is depicted in Figure~\ref{tab:dataset-structure}.

Each JSON object has the following three main fields:

\begin{enumerate}
  \item 
  % The first field is 
  \texttt{\_id}~--~the unique identifier of each entry
  \item 
  % The second field is the 
  \texttt{citing\_paper}~--~an object holding the \texttt{dblp\_id} of each citing paper
  \item 
  % The third field is called 
  \texttt{cited\_papers}~--~
  % and is 
  an array that contains the objects that correspond to each reference found in the text of the \texttt{citing\_paper}. Each object of the array may contain some or all of the following fields: 
  
  \begin{enumerate}
      \item \texttt{dblp\_id}~--~the dblp\_id of the cited paper
      \item \texttt{doi}~--~the doi of the cited paper
      \item \texttt{bibliographic\_reference}~--~the raw citation string as it appears in the citing paper
  \end{enumerate}
\end{enumerate}

Note that not all the aforementioned fields in ($3$) are required for a \texttt{cited\_paper} to be valid.
Specifically,
one of the \texttt{dblp\_id} or \texttt{doi} identifiers is required for a cited paper to be added in the collection. 
Finally, the \texttt{bibliographic\_reference} exists in all cited\_paper objects since it is extracted directly from the PDF files of each citing paper in the dataset.

Table~\ref{tab:stats}~summarises some statistics about the BIP! NDR dataset. In particular, $59,663$ full texts from Open Access papers were parsed. A total of $1,054,107$ references were evaluated, and among them, $511,842$ references were successfully matched with corresponding keys from the DBLP database. Additionally, $366,106$ DOIs were successfully matched with these DBLP keys. Finally, there were $22,569$ DOIs that could not be matched with any DBLP key, indicating that they have not been indexed by DBLP. 

\begin{table}
    \caption{Statistics of BIP! NDR dataset (current version).}\label{tab:stats}
    \centering
    \begin{tabular}{|l|l|l|}
        \hline
         Statistic & \#  \\ 
        \hline
        Total Files Parsed & 59,663 \\
        Total References Evaluated & 1,054,107 \\
        DBLP Keys Matched & 511,842 \\
        DOIs Matched with DBLP Key & 366,106 \\
        DOIs without DBLP Key & 22,569 \\
        \hline
    \end{tabular}
\end{table}

%% file: sections/sec5-conclusions.tex
\section{Conclusions}\label{sec5-conclusions}

We presented BIP! NDR, a dataset created using text analysis techniques
on the DBLP database to extract citation information from the full text of the Open Access papers that do not have an assigned DOI. The dataset 
offers over 500K citations from Computer Science papers that do not have DOIs, addressing a significant limitation of widely used citation datasets in the field, that fail to cover them. As a result, it enables more comprehensive and accurate research assessment in Computer Science. In the future, we plan to improve the workflow so that it can identify more Open Source publications and to extend the dataset so that it can offer additional metadata for each citation (e.g., a class according to a citation classification algorithm). 

\section*{Acknowledgements}
This work was co-funded by the EU Horizon Europe projects SciLake (GA: 101058573) and GraspOS (GA: 101095129).

\begin{figure}[!h]
     \centering
         \includegraphics[width=0.1\linewidth]{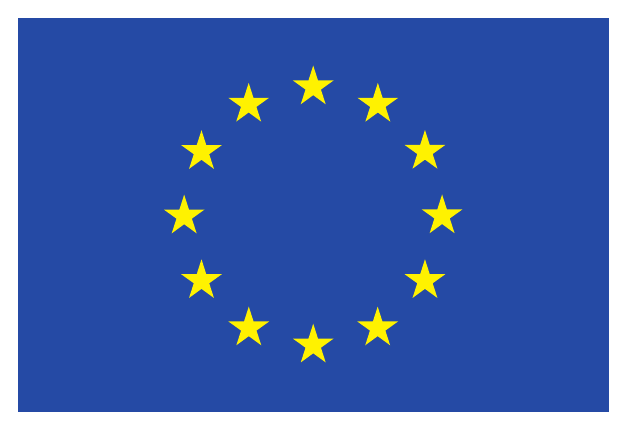}
\end{figure}

%% file: main.bbl
\begin{thebibliography}{}

\bibitem[Abu-Jbara et~al., 2013]{AER2013}
Abu-Jbara, A., Ezra, J., and Radev, D. (2013).
\newblock Purpose and polarity of citation: Towards nlp-based bibliometrics.
\newblock In {\em Proceedings of the 2013 conference of the North American chapter of the association for computational linguistics: Human language technologies}, pages 596--606.

\bibitem[Färber and Ao, 2022]{mag-qss}
Färber, M. and Ao, L. (2022).
\newblock {The Microsoft Academic Knowledge Graph enhanced: Author name disambiguation, publication classification, and embeddings}.
\newblock {\em Quantitative Science Studies}, 3(1):51--98.

\bibitem[Hendricks et~al., 2020]{crossref}
Hendricks, G., Tkaczyk, D., Lin, J., and Feeney, P. (2020).
\newblock {Crossref: The sustainable source of community-owned scholarly metadata}.
\newblock {\em Quantitative Science Studies}, 1(1):414--427.

\bibitem[Koloveas et~al., 2023]{Koloveas-bipndr}
Koloveas, P., Chatzopoulos, S., Tryfonopoulos, C., and Vergoulis, T. (2023).
\newblock Bip! ndr (nodoirefs): a dataset of citations from papers without dois in computer science conferences and workshops.

\bibitem[Ley, 2002]{ley2002dblp}
Ley, M. (2002).
\newblock The dblp computer science bibliography: Evolution, research issues, perspectives.
\newblock In {\em String Processing and Information Retrieval: 9th International Symposium, SPIRE 2002 Lisbon, Portugal, September 11--13, 2002 Proceedings 9}, pages 1--10. Springer.

\bibitem[Ley, 2009]{ley2009dblp}
Ley, M. (2009).
\newblock Dblp: some lessons learned.
\newblock {\em Proceedings of the VLDB Endowment}, 2(2):1493--1500.

\bibitem[Lo et~al., 2020]{LWN2020}
Lo, K., Wang, L.~L., Neumann, M., Kinney, R.~M., and Weld, D.~S. (2020).
\newblock S2orc: The semantic scholar open research corpus.
\newblock In {\em ACL}.

\bibitem[Lopez, 2009]{L2009}
Lopez, P. (2009).
\newblock Grobid: Combining automatic bibliographic data recognition and term extraction for scholarship publications.
\newblock In {\em International conference on theory and practice of digital libraries}, pages 473--474. Springer.

\bibitem[Meuschke et~al., 2023]{MJS2023}
Meuschke, N., Jagdale, A., Spinde, T., Mitrovi{\'c}, J., and Gipp, B. (2023).
\newblock A benchmark of pdf information extraction tools using a multi-task and multi-domain evaluation framework for academic documents.
\newblock In Sserwanga, I., Goulding, A., Moulaison-Sandy, H., Du, J.~T., Soares, A.~L., Hessami, V., and Frank, R.~D., editors, {\em Information for a Better World: Normality, Virtuality, Physicality, Inclusivity}, pages 383--405, Cham. Springer Nature Switzerland.

\bibitem[Nicholson et~al., 2021]{NML2021}
Nicholson, J.~M., Mordaunt, M., Lopez, P., Uppala, A., Rosati, D., Rodrigues, N.~P., Grabitz, P., and Rife, S.~C. (2021).
\newblock {scite: A smart citation index that displays the context of citations and classifies their intent using deep learning}.
\newblock {\em Quantitative Science Studies}, pages 1--17.

\bibitem[Papastefanatos et~al., 2020]{observatory}
Papastefanatos, G., Papadopoulou, E., Meimaris, M., Lempesis, A., Martziou, S., Manghi, P., and Manola, N. (2020).
\newblock Open science observatory: monitoring open science in europe.
\newblock In {\em ADBIS, TPDL and EDA 2020 Common Workshops and Doctoral Consortium: International Workshops: DOING, MADEISD, SKG, BBIGAP, SIMPDA, AIMinScience 2020 and Doctoral Consortium, Lyon, France, August 25--27, 2020, Proceedings 24}, pages 341--346. Springer.

\bibitem[Peroni and Shotton, 2020]{peroni-qss}
Peroni, S. and Shotton, D.~M. (2020).
\newblock Opencitations, an infrastructure organization for open scholarship.
\newblock {\em Quant. Sci. Stud.}, 1(1):428--444.

\bibitem[{The DBLP Team}, 2023]{dblp-dataset}
{The DBLP Team} (2023).
\newblock dblp computer science bibliography. monthly snapshot release of november 2022.

\bibitem[Tkaczyk et~al., 2018]{TCS2018}
Tkaczyk, D., Collins, A., Sheridan, P., and Beel, J. (2018).
\newblock Machine learning vs. rules and out-of-the-box vs. retrained: An evaluation of open-source bibliographic reference and citation parsers.
\newblock In {\em Proceedings of the 18th ACM/IEEE on joint conference on digital libraries}, pages 99--108.

\bibitem[Vergoulis et~al., 2022]{bip-finder}
Vergoulis, T., Chatzopoulos, S., Vichos, K., Kanellos, I., Mannocci, A., Manola, N., and Manghi, P. (2022).
\newblock Bip! scholar: a service to facilitate fair researcher assessment.
\newblock In Aizawa, A., Mandl, T., Carevic, Z., Hinze, A., Mayr, P., and Schaer, P., editors, {\em {JCDL} '22: The {ACM/IEEE} Joint Conference on Digital Libraries in 2022, Cologne, Germany, June 20 - 24, 2022}, page~42. {ACM}.

\bibitem[Vergoulis et~al., 2021]{bipdb}
Vergoulis, T., Kanellos, I., Atzori, C., Mannocci, A., Chatzopoulos, S., Bruzzo, S.~L., Manola, N., and Manghi, P. (2021).
\newblock Bip! {DB:} {A} dataset of impact measures for scientific publications.
\newblock In Leskovec, J., Grobelnik, M., Najork, M., Tang, J., and Zia, L., editors, {\em Companion of The Web Conference 2021, Virtual Event / Ljubljana, Slovenia, April 19-23, 2021}, pages 456--460. {ACM} / {IW3C2}.

\bibitem[Yousif et~al., 2019]{YTN2017}
Yousif, A., Niu, Z., Tarus, J.~K., and Ahmad, A. (2019).
\newblock A survey on sentiment analysis of scientific citations.
\newblock {\em Artificial Intelligence Review}, 52(3):1805--1838.

\end{thebibliography}
